\documentclass[structabstract]{aa} 
\usepackage{graphicx}
\usepackage{float}
\usepackage{txfonts}
\usepackage{natbib}
\usepackage{subfig}

\newcommand{\lsp}{LS~I~+61$^{\circ}$303}
\newcommand{\lsi}{LS~I~+61$^{\circ}$303~}

\newcommand{\beq}{\begin{equation}}
\newcommand{\eneq}{\end{equation}}

\begin{document}

\title{
VLBA images of the precessing jet of LS I +61$^{\circ}$303}

\author{M.  Massi 
        \inst{1},
E. Ros
  \inst{2,1},
    \and
        L.\ Zimmermann
         \inst{1}
}

\institute{
 Max-Planck-Institut f\"ur Radioastronomie, Auf dem H\"ugel 69, D-53121 Bonn, Germany \\
\email{mmassi@mpifr-bonn.mpg.de,  lzimmerm@mpifr-bonn.mpg.de}
 \and
Departament d'Astronomia i Astrof\'{\i}sica, Univ. de Val\`encia, E-46100 Burjassot,
Valencia, Spain \\
\email{Eduardo.Ros@uv.es}}

   \date{Draft: \today}

\abstract
{
In 2004, changes in the radio morphology of the Be/X-ray binary 
system \object{LS~I~+61$^{\circ}$303~} suggested that it is a precessing microquasar.
In 2006, a set of VLBA observations performed throughout the entire  orbit
of the system  were not used to study its precession because the  changes in
radio morphology could tentatively be explained by  the alternative  pulsar  model.
However, a recent  radio spectral index data analysis has  
confirmed the predictions of the two-peak microquasar model, which therefore does apply 
in \lsp.} 
{We revisit the set of VLBA observations performed throughout  the orbit to  determine the precession  period and 
improve our  understanding of the physical mechanism behind the precession.}
{By  reanalyzing the  VLBA data set, we improve the dynamic range of images by a factor of four, using self-calibration. Different fitting techniques are used and 
compared to determine 
the peak positions in phase-referenced maps.}
{The  improved dynamic range shows that in addition to the images with a one-sided structure,
 there are  several images 
with a double-sided structure. 
The astrometry indicates that the peak in consecutive images for the whole set of observations
 describes a well-defined  ellipse,  6$-$7  times larger than the  orbit, with  a period of about 28~d.}
{A double-sided structure is  not 
expected to be formed from the  expanding  shocked wind predicted in the pulsar scenario. 
In contrast,  a precessing   microquasar model  can explain the double- and one-sided
structures in terms of variable Doppler boosting.
The ellipse defined by the astrometry could be the cross-section of the
precession cone, at the distance
of the 8.4\,GHz-core of the steady jet,
and  $~$28\,d the precession period.}


\titlerunning{VLBA images of \lsi}

\maketitle

\begin{figure*}
 \centering
\includegraphics[width=.75\textheight]{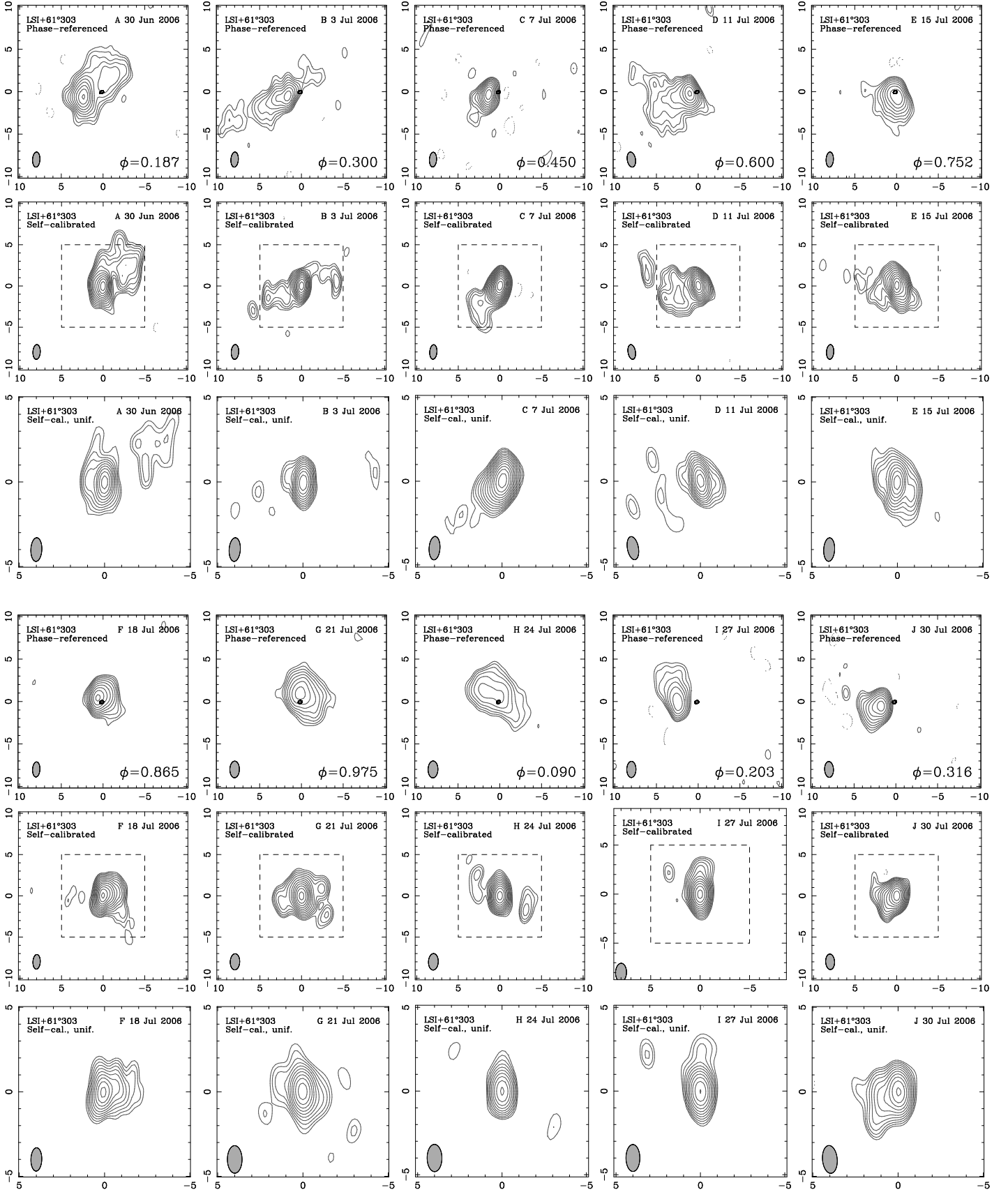}
      \caption{
Images of VLBA  runs A-J at 8.4\,GHz (3.6\,cm) of \lsp.
The units on the axes are milliarcseconds (mas).
For each run, 
three  maps  are presented, the phase-referenced map 
 (beam of 1.8 - 2.0  mas $\times$ 0.9 - 1.1  mas, shown in the bottom left
corner), the 
self-calibrated map at the same resolution, and the self-calibrated  map 
with a beam of 1.4 - 1.6  mas $\times$ 0.6 - 0.9   mas. 
Contour levels for all maps are -4, 4, 5.66, 8, 11.3, 16, 22.6, 32, 45.2,
64, 90.5, 128, 181 $\sigma $ 
(with 1$\sigma$=0.1 mJy/beam for the  top and bottom maps,  1$\sigma\simeq$0.07
middle maps).
At the center of the  phase-referenced images, 
we trace, in scale,  the orbit (see Sect. 2).
      }
         \label{maps}
   \end{figure*}

\section{Introduction}
In 1993,  
a VLBI observation of \lsi showed that the radio emission had a structure of milliarcsecond
(mas) size corresponding to a few AU at the distance of 2.0~kpc 
\citep{massi93, frailhjellming91}. 
The complex
morphology in successive VLBI observations 
\citep{peracaula98,paredes98,taylor00,massi01,massi04}
highlighted the unlikelihood of  an interpretation in
terms of a microquasar  with a constant position angle.
The radio morphology  not only changes  position angle, but  it is even sometimes one-sided 
and at other times 
 two-sided. This suggested the hypothesis of \lsi being a precessing microblazar 
\citep{kaufman02,massi04}.
A microblazar,
i.e., a microquasar with
radio jets forming a small angle, $\theta$, with respect to the observer's line of sight, 
 has been proposed to be the Galactic version of  extragalactic blazars. 
Doppler boosting enhances the radiation from material that is moving towards the observer, 
 and attenuates it when it moves in the opposite direction.
In cases of substantial  flux attenuation in the receding jet 
(i.e., attenuation to a level fainter than the sensitivity of radio images), the structure will
appear as a one-sided jet. 
A precession of the jet leads to a variation in the  angle, $\theta$,
and therefore variable Doppler boosting. The result is both a continuous 
variation in  the position angle of the radio-emitting structure
and of the flux density  ratio of the approaching to the receding jet 
 \citep{massi07}.

In  known precessing X-ray binaries, 
the timescale for tidally forced precession of the accretion disk around the compact object, induced by the companion star, lies within the range
$8-22$\,times the orbital period \citep{larwood98, massizimmermann10}.
In this context, the peculiarity of the  variations in \lsi is their short timescale
 with respect to the orbital period of 26.496~d.
 \citet{massi04} found that  MERLIN images revealed a surprising variation
of 60$\degr$ in position angle in only one day.
Even if quantitatively the relationship between position angle in the image 
and the viewing angle $\theta$
is not straightforward,  a rapid  variation in the  position angle implies nevertheless 
that there is clearly a rapid variation in $\theta$.
The fast position angle variation has been confirmed by VLBA observations.
\citet{dhawan06}  measured in  VLBA images a rotation of the inner structure of
roughly $5\degr-7\degr$ in 2.5 hrs, that is again a variation of almost 60${\degr}$/day,
but suggested that the variations were  due to 
a cometary tail of a pulsar.
If the compact object is a pulsar,  the interaction between its   relativistic wind  and
the   equatorial wind of the Be star  is predicted to create
a bow-shock around the pulsar  with a sort of cometary tail, i.e., a one-sided structure,  extending  away from   
  the Be star   \citep{dubus06}. 
However, the  analysis of the radio spectral index by \citet{massikaufman09}
prove that  \lsi displays the typical characteristic of   microquasars
of an optically thin outburst after  an interval of  optically thick emission.
In microquasars,  the so-called transient jet associated with the large optically thin outburst, 
is related to shocks travelling in a pre-existing  steady jet, 
that is a slow-moving continuous conical outflow with  a composite flat/inverted radio spectrum
(i.e., optically thick emission) \citep{fender04, massi11}.
The remarkable finding for  \lsi is that,
during the maximum of its long-term periodicity (4.6 yr) 
the alternance between  optically thick and  optically thin emission,
occurs twice  during the orbit, 
first around periastron and then again, shifted by almost 0.3-0.4 in orbital phase 
(0.3-0.4 $\times$ 26.496~d = 8-11~d after periastron), i.e. towards apastron \citep{massikaufman09}. 
This agrees  with 
 the well-known ``two peak accretion/ejection model'', 
which has been applied  by several authors to  \lsp, 
predicting,  for large mass accretion rate,  $\dot{M}$,  two events: one around periastron and the second shifted about 0.3 in orbital phase towards apastron
\citep{taylor92, martiparedes95, boschramon06, romero07}.

The radio spectral index data corroborate the microquasar model for \lsp.
The most important uncertainty is therefore the process that could  produce the observed fast precession.
An essential step in any investigation is to establish the precessional period.
This parameter could  be derived by the  reanalysis of the  VLBA observations. These observations
were  performed by  
Dhawan et al.\ (2006) every 3 days over  30 consecutive days, towards the minimum of  the long-term periodicity. 
\citet{dhawan06} suggested that the  peaks of the maps  
trace an erratic  ellipse.
In a precessing microblazar, as explained below,
the core component of the steady jet  describes an ellipse, whereas the transient jet
adds  random shifts.   
To determine important precession parameters, e.g. the period, we therefore reanalyzed the
set of VLBA observations performed  by \citet{dhawan06}.
\section{Data reduction and results}
\lsi was observed by the VLBA on ten  different days from 2006 June 30 (segment A) to 2006 July 30 (segment J).  The observations (code BD117) included data at 2.3\,GHz 
and 8.4\,GHz.
We performed the post-correlation data reduction using 
AIPS.  The phase and amplitude calibration were performed in a
standard way: corrections for residual Earth orientation parameters 
in the correlator model
and ionosphere---in the latter case those derived from GPS data---were introduced; the digital 
sampling amplitude corrections (ACCOR) were applied; 
system temperature and gains were applied to get the amplitude
calibration (APCAL); delay offsets between sub-bands
were corrected using pulse-tones measured during the observations (APCOR);  a parallactic angle correction was introduced;
and the final delay and rate calibration across the observing band was 
performed using the  FRING  routine.  
After these procedures, the data
for the phase reference calibrator, \object{J0244+6228}, were exported to be
imaged in Difmap.  We read the image back into AIPS 
and calibrated the amplitudes and phases for the source and 
produced a hybrid map of the source with IMAGR  within AIPS, 
that matched the results obtained with Difmap.  After that,
the delay and rate were calibrated again with FRING after
dividing the data by the clean-component model produced
by IMAGR  for this source
to get structure-free solutions for J0244+6228.  The structure-free
solutions obtained were then interpolated to the target source, \lsp, and those 
data were exported to be imaged.
\subsection{Analysis and results of the 8.4\,GHz data}
The 8.4\,GHz  phase-referenced images  
(with respect to  J0244$+$6228)
shown in Fig.~\ref{maps},
closely reproduce the Dhawan et al.\ (2006) structures.
At their center,  we show the orbit with a semimajor axis of 
$(0.22-0.23)$\,mas ($\simeq 0.7\times 10^{13}$ cm at a distance of 2\,kpc), 
that was derived from the third  Keplerian law
for $P=26.496$\,d with a mass for the  B0 star of 17\,$M_{\odot}$ and a mass for the compact object of $1.4 M_{\odot} - 4 M_{\odot}$, and traced using \citet{casares05}  orbital parameters. 
The orbit  is  not only much smaller than the radio-emitting region,
but also smaller  than  the interferometric  beam of the VLBA at this wavelength. 
We make this remark  because  the zoomed orbit in  Fig.~3 by Dhawan and collaborators
(zoomed to  show more clearly the orbital phase of each run) 
may unfortunately produce  the wrong  impression that the orbit is resolved
    when  it is not, and only the accuracy of the astrometry, which is 
      0.04\,mas at 8.4\,GHz, is of a smaller scale  than the orbit (see Sect.\,2.2). 

\begin{figure}
 \centering
\includegraphics[width=.43\textwidth, clip, angle=0]{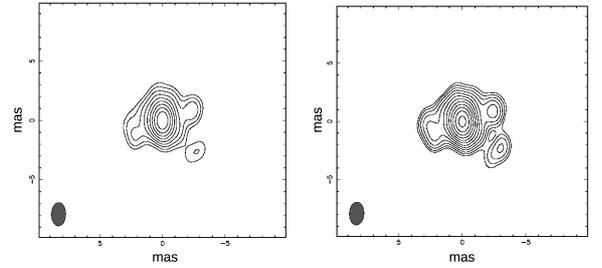}\\
\caption{
Left: 
Intermediate  self-calibrated image of run G,  at 8.4 GHz, after eight manual iterations of hybrid mapping with
phase-only self-calibration. 
The beam is  1.96 mas $\times$ 1.14 mas at -2\degr.
           Contour levels  are -4, 4, 5.66, 8, 11.3, 16, 22.6, 32, 45.2,
           64 $\sigma $, with 1$\sigma$=0.18 mJy/beam.
Right:  Output of  automatic hybrid mapping by  Muppet. 
Muppet switches on amplitude plus phase self-calibration
after phase-only self-calibration cycles have converged.
The beam is 1.96 mas $\times$ 1.14 mas at 1 \degr.
                   Contour levels   are -4, 4, 5.66, 8, 11.3, 16, 22.6, 32, 45.2, 
                   64, 90.5, 128, 181 $\sigma $, 
                   with 1$\sigma$=0.08  mJy/beam.
}
\label{aipsG}
\end{figure}

For each run, we present in Fig.~\ref{maps} both the  phase-referenced  
and  the  self-calibrated images.
Self-calibrated images were produced
with the automatic procedure Muppet  of the Caltech Difmap imaging package  \citep{shepard97}.
Using self-calibration, the information on absolute flux density and absolute position are lost,
but the removal of residual calibration errors improves the dynamic range
by up to a factor of four  for our maps \citep{cornwellfomalont99}.
Muppet starts with a point source as an initial model
then automatically selects the clean boxes around peaks with 
signal-to-noise ratio (S/N) values $\geq$\,6,
and switches on amplitude plus phase self-calibration
after phase-only self-calibration cycles have converged.
The high limit of the signal-to-noise ratio S/N\,$\geq$\,6 prevents 
the risk of creating artifacts, which are only introduced when 
self-calibrating  below  S/N\,$<$\,4   \citep{martividal08}.
For epoch G, there is a significant difference between the phase-referenced
image and the self-calibrated image at the same resolution (see Fig.  \ref{maps}).
Therefore, for this epoch, we also self-calibrated the data  manually. 
Figure \ref{aipsG} shows an intermediate  image after eight manual iterations of hybrid mapping starting
 with a point source and  using  phase-only self-calibration.  
The two images of Fig.~\ref{aipsG} clearly show  the  same  structures only 
with different flux densities. This difference 
depends upon  further iterations of 
amplitude self-calibration performed by Muppet.

\begin{figure}
 \centering
\includegraphics[width=.47\textwidth, clip, angle=0]{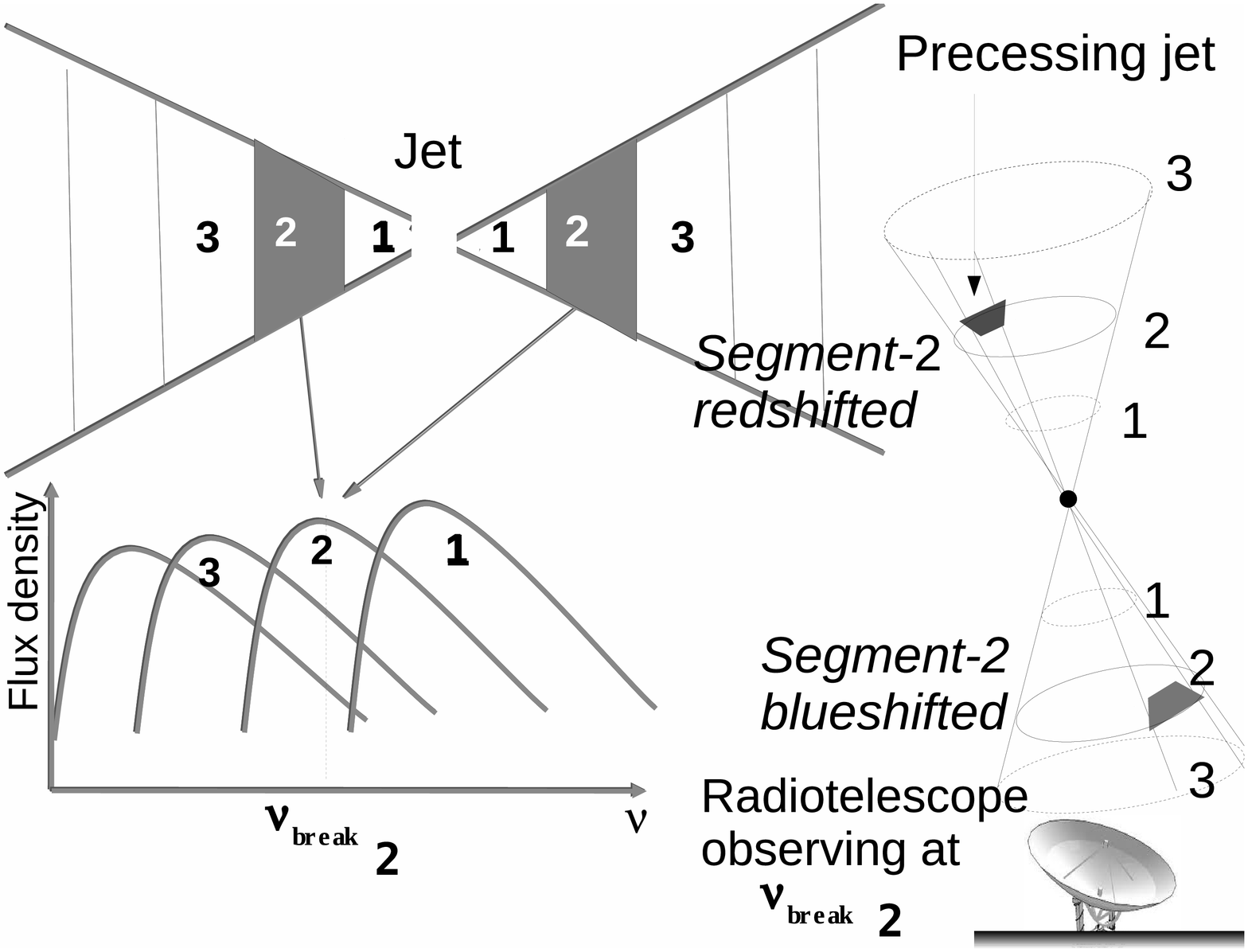}\\
\includegraphics[width=.35\textwidth, clip, angle=-90]{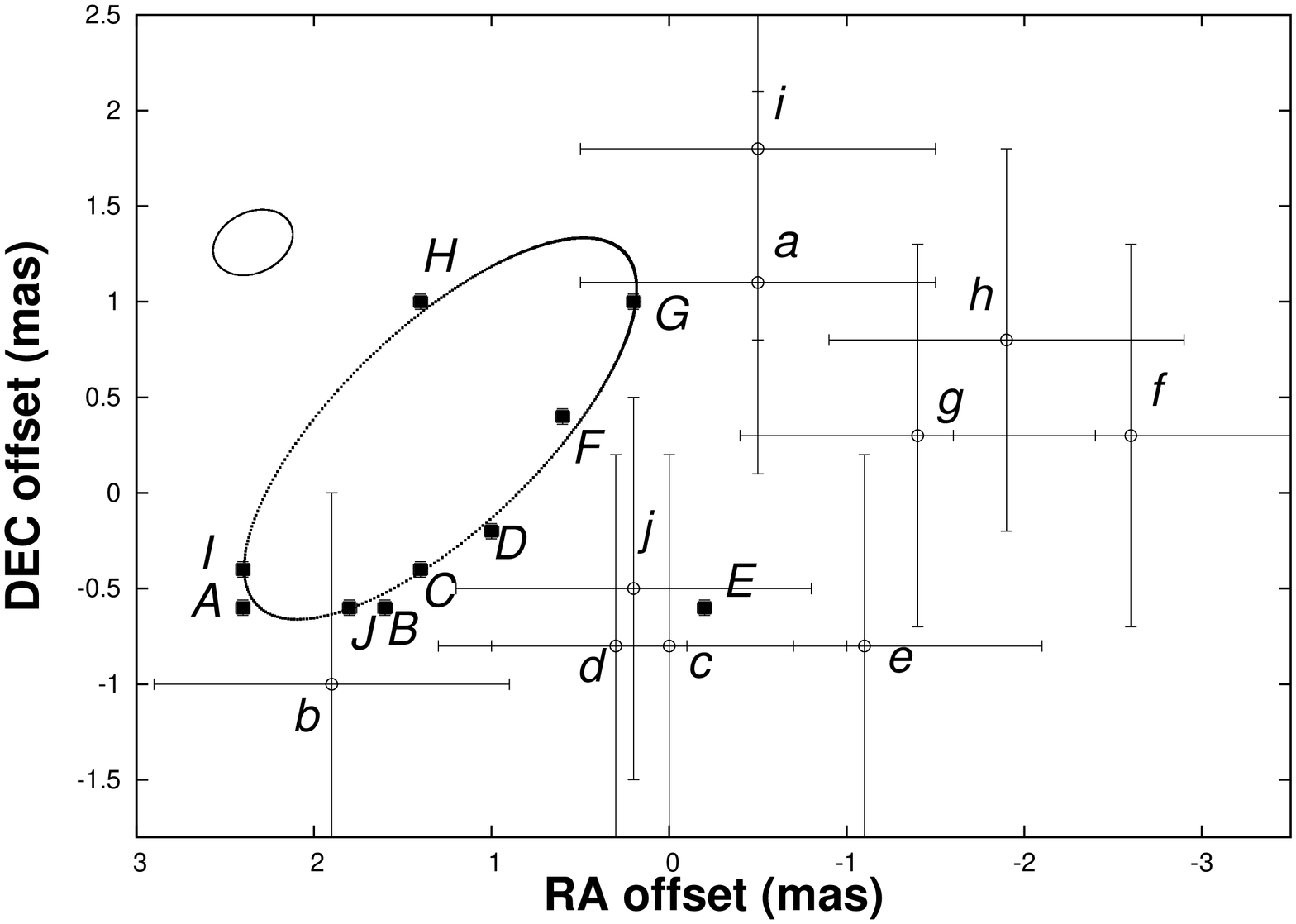}\\
\caption{
Top-left: Superposition of individual spectra, 
each with a different $\nu_\mathrm{break}$,
associated with different  segments of a steady jet.
Top-right: For a precessing microblazar,
the  core component,  dominated by the approaching jet contribution because of Doppler boosting, will describe 
an ellipse during  precession. 
Bottom: Astrometry of consecutive peaks of  VLBA  8.4\,GHz maps (capital letters)
and 2.3\,GHz maps (small letters)
for runs A-J
with the orbit drawn to scale at an arbitrary distance. East is to the left.
Peak of 8.4\,GHz map E, even if displaced from the other ones,
 which is therefore likely affected by the approaching component of the
transient jet,
is at a position angle that is consistent  with  the other peaks.
The small error of 0.04 mas for the astrometry at 8.4 GHz is indicated by the filled squares.
That the precessing compact object  moves in an orbit with
semimajor axis of
  $\sim$0.2\,mas introduces additional variations.
Errorbars  of 1 mas  for the astrometry at  2.3\,GHz represent only a lower limit to the error 
associated with large Galactic scattering and low resolution (see Sect. 2.2).}
 \label{sketch}
\end{figure}
The determination of the peak position in the phase-referenced
maps is an issue, especially if the source is not Gaussian-like
or symmetric, as  was the case in some epochs.  The first approach
is to determine the location of the brightest pixel with IMSTAT, then 
three different methods were applied to fit the  position 
either in the image plane with JMFIT   and  MAXFIT, 
or in the visibility plane, with modelfit   in Difmap.
All results from the four applied methods coincide within $(1-3)\,\sigma$,
trace an ellipse over time, and  show that the peak of run E 
is displaced from the other peaks.
However, the Gaussian fits have asymmetric residuals, 
which  indicates  that the brightest parts of the images are  asymmetric.
As discussed in Sect. 3, the
core of the steady jet for a microblazar is clearly not Gaussian,
because of Doppler boosting: the redshifted part should be fitted
with a smaller half ellipse  than the blueshifted part.
For consistency, we use for 
Fig.~\ref{sketch}  the results for IMSTAT
(coincident within 1\,$\sigma$ with MAXFIT). 
It is evident that the  peaks of the consecutive images clearly trace  an ellipse 
6$-$7  times larger than the  orbit.
Galactic scattering at  8.4\,GHz, as discussed in the next section, gives rise to an  error of  0.04 mas.
This value is slightly smaller  than the  average value of 0.05 mas 
estimated for  
the astrometric accuracy of
a full track phase-referenced VLBI experiment \citep{pradel06}.

In the images, the larger dynamic range of the self-calibrated maps show
in the image of run A a double-component structure. 
This  is clearly unlikely to correspond to a pulsar nebula,
whereas it could be compatible with either the approaching and receding component of a transient jet,
or with variable Doppler boosting along a one-sided twisted jet. 
Image B  
already shows in the phase-referenced map with natural weighting a double-sided morphology difficult to reconcile
with the pulsar nebula. The self-calibrated map shows 
an intriguing  twisted jet as expected for a fast precession. 
Runs C, F, I, and J show a one-sided jet or alternatively a pulsar cometary-tail but again  
runs D, E, G,  and H show the double-sided jet structure.

\subsection{Analysis and results of the 2.3\,GHz data}
In Fig.~4 of \citet{dhawan06}, the   2.3\,GHz astrometry  
seems to show a shift of up 2 mas from  the  8.4\,GHz astrometry.
In practice, no hypothesis can be made lacking the  points of  error bar.
We show in this section that,  whereas the accuracy  associated with the peaks at  8.4\,GHz   is 0.04 mas,
that associated with the peaks at 2.3 GHz is more than one order of magnitude higher.

We re-examined  the  2.3\,GHz data.   
Large amounts of Galactic scattering at this frequency affects the data, broadening the images.  This effect is  discussed  in detail in the astrometric works of Lestrade
 and collaborators \citep{lestrade85, lestrade99},
who  attributed  the   size of
4.1 mas,  determined at  1.62\,GHz, to broadening caused by  interstellar scattering.
Therefore, at 2.3\,GHz the broadening,  which is  proportional to $\lambda^2$, is  2.2 mas
(whereas at 8.4\,GHz, it is only 0.16 mas).
The additional complicating factor is that  refractive interstellar scintillation will cause the apparent
source position to wander, and   the magnitude of this wandering is hard to calculate 
(see, e.g. \citet{rickett90}).

\begin{figure}
\centering
\includegraphics[width=.35\textwidth,  angle=-90.]{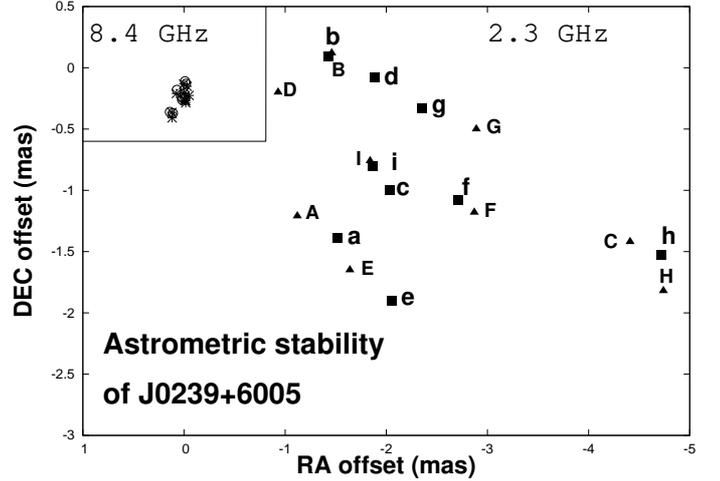}
\caption{
Control astrometry on the secondary calibrator.  Relative positions of
 J0239+6005  with respect to  J0244+6228, 2.5\degr apart.
The astrometry at 2.3\,GHz is indicated by squares (JMFIT) and triangles
(MAXFIT); that at 
 8.4\,GHz is indicated by  crosses (JMFIT) and
circles (MAXFIT).
}
         \label{j0239}
\end{figure}

We were able to estimate  the quality of the astrometry  from the scatter in positions
of the check source observed in project BD117, namely \object{J0239+6005},
that is  2.5\degr apart from the calibrator J0244+6228. 
We determined the astrometric stability of that check source at 8.4\,GHz and 2.3\,GHz
and used the results to assign error bars to the astrometric measurements at both frequencies.
First, we performed the phase-referencing analysis of the source J0239+6005
with respect to the phase reference calibrator J0244+6228, and then  determined the peak position
in the maps of J0239+6005  for the consecutive runs.
As shown in Fig.~\ref{j0239},  when there is no precession and one expects all results to  coincide
 within their errors, the  points at 8.4\,GHz  indeed overlap 
and have  a dispersion of 
 0.08 mas.  
The result at low frequency is completely different. The scatter is huge: 
the dispersion in the points at 2.3 GHz is  1.2 mas. 
The  astrometric error scales with separation from the phase reference calibrator \citep{pradel06}. 
 \object{J0239+6005} is 
 2.5\degr apart from the calibrator J0244+6228, whereas 
\lsi is 1.4\degr away from  J0244+6228. 
Therefore, the 0.08 mas rms  for 8.4 GHz data 
and the  1.2 mas rms for 2.3 GHz data  of J0239+6005  scales  to   
0.04 mas  and 0.7 mas, respectively,  for \lsp. 
Moreover, at 2.3 GHz, 
 besides the already large error  
of 0.7 mas 
 associated with the  scattering, there is also
the error due to the low resolution.
In fact, at 2.3 GHz, \lsi  is  unresolved/barely resolved
  with the VLBA (see  Fig.~5).
  In the large beam,   4 mas $\times$ 7 mas,
  the optically thin emission of the extended jet may contribute substantially
  and, depending on its brightness, displace the  peak out of the core position, especially in the case of
  one-sided structures.
 At some low level, we can test this effect with the data at 8.4\,GHz, in a similar way
  as done in Fig. 4 by Dhawan et al. (2006) where one can see  peak and
  "centroid" positions for 3.6 cm (i.e. 8.4\,GHz). We strongly tapered the
  data at 8.4\,GHz to reproduce
  the large beam at 2.3\,GHz. We then fitted the peak positions in the
  low resolution maps obtained 
  and  compared them  with those fitted in  the maps of Fig. 1 (i.e.  the
  points along the
  ellipse of Fig. 3). 
This test at  8.4\,GHz, which can obviously not take into account resolved larger structures, 
measures  differences between  positions 
with a dispersion of  0.6 mas.
This error adds quadratically to the error of about 0.7 mas associated with the scattering,  
resulting in  the  total error of $\sim$1 mas for the astrometry of \lsi  given in Fig. 3-bottom. 
We note that  with this  value, which represents, however, only a lower limit to the error bars, there is an overlap
 at each epoch at  2.3\,GHz of the results from the different fitting programs,
which can differ by  up  to 2 mas.

We  note that the  2.3\,GHz astrometry has a general shift with
respect to that at
 8.4\,GHz. This, as discussed in the next section, would correspond to the predictions of
 the core-shift effect.
 Owing to this effect, the peak of the emission
(i.e. core) representing the optically thick part of the jet at one  observing
 frequency,  is predicted to have a shift from the center that is different  for the different frequencies (Fig. 3-Top).
 This  shift can also be seen  in the  2.3\,GHz
 astrometry of  Dhawan et al. (2006) without the  error bars. 
As in our case,  Dhawan et al.  ascribed the shift to a synchrotron opacity  gradient  along the emitting structure.
However, in that case the structure is not thought to be a precessing jet but  a cometary tail. 
Anything  more than a general trend cannot be inferred from the  2.3\,GHz data.
As a result of our analysis of the data  at  2.3\,GHz, it is indeed clear that any quantitative comparison
between the astrometry at 8.4\,GHz and at  2.3\,GHz
(as well as   between the orientation of the radio structures at the two frequencies)
is unfortunately  prevented by the  very large Galactic  scattering and  
the low resolution.

	\section{Precessing microblazar}

Figure~\ref{sketch}-bottom shows the results of the astrometry for the $\nu$=~8.4\,GHz data: the  peaks of the consecutive images 
clearly trace an ellipse. 
Here, we show that this  ellipse could be  the cross-section of the 
precession cone, at the distance
of the 3.6 cm-core  ($\nu$= 8.4\,GHz) of the steady jet.

A magnetized plasma containing energetic electrons with a power-law energy distribution
will produce a synchrotron power-law spectrum.  However, below a critical frequency
($\nu_\mathrm{break}$),
the radiating electrons will re-absorb some of the photons and as a result the
typical  spectrum
of a uniform synchrotron source will show a peak at $\nu_\mathrm{break}$.
Changes in the electron energy distribution and the
decay of the magnetic field
along the  conical outflow forming the steady jet,  imply that the critical  frequency varies along the jet \citep{marscher95, massi11}.

In microquasars, the  $\nu_\mathrm{break}$ for the part of the steady jet closest to the engine (i.e.,
 $\nu_\mathrm{break_1}$ in Fig.~3) lies in the infrared \citep{russell06},
       whereas in AGNs the observed turnover 
lies in the millimeter range \citep{marscher95}.
Observing at longer wavelengths than the infrared, i.e. in the  radio band, gives
rise to two results: a flat spectrum and the ``core shift''.
Multi-wavelength observations result  in the flat/inverted spectrum discussed in the introduction,
which is typical of a steady jet.
Imaging a steady jet at one observing frequency,  $\nu_\mathrm{obs}$, 
gives rise to the effect known as ``core shift'', where the displacement  from the center  is a
function of the observing frequency  $\nu_\mathrm{obs}$.
At $\nu_\mathrm{obs}$, the emission of the segment will dominate, whose spectrum peaks at
that frequency
plus small contributions from neighboring segments (see Fig.~1 in \citet{markoff10}).
With the engine being  close to the ``infrared-core", it is clear that  
the 3.6~cm-core can be  displaced away from the orbit (as  segment 2 is in  Fig.~3 top-left).
If a jet  is pointing  towards us (i.e., a microblazar as in Fig.~3 top-right),  
the core will be dominated by 
the approaching jet,  because of Doppler boosting.
If the jet is precessing, then the core will  describe    
an  ellipse (see  Fig.~3 top-right).
A transient jet, in contrast, can undergo any shift from the center, depending on the velocity
and the  time elapsed from the transient.
The peak of map E, even if displaced away from the others,
 and therefore likely affected by the approaching component of the transient jet, is at a 
position angle which is consistent  that  the other peaks. 
 After 27~d, the peak of run I is  rather close to completing the  cycle, i.e. to
overlapping  with the peak of the starting run A. The same occurs for the peak of run J, that is 
27 d after run B, a peak that
nearly overlaps  with peak B. 
The peak of run J, 30 d after run A, is clearly 
displaced from the A peak. The period  therefore seems to be in the range 27-28 d.

\section{Conclusions and discussion}
       \citet{dhawan06} suggested that  their  VLBA observations of \lsi 
could probe  the cometary tail of the pulsar model and  that the peaks  of all images  traced
       an erratic ellipse.
In this paper, we have re-analyzed this data set and have shown that
\begin{enumerate}
\item
The larger dynamic range of the self-calibrated maps reveals that the radio emission has 
in several images a double-sided structure. 
\item
Our fit analysis show that the peaks of the images at 8.4\,GHz trace a well-defined  ellipse 
over a time interval of about 27-28\,d. 
\end{enumerate}

The  pulsar model explains neither the double-sided morphology nor the change 
from a double-sided to a  one-sided structure.
The  microquasar model can explain them in terms of  variable Doppler boosting, i.e., with a  precessing jet.
The cm-core of a  precessing steady jet pointing close to our line of sight, 
as in a microblazar,
is  expected to describe an ellipse during the precession.  In the transient jet phase,
there will be an additional   shift due to  the approaching jet component.   
We conclude therefore that the precession period is the time of about 27-28\,d 
necessary to
complete the ellipse.  

As shown in Fig.~\ref{sketch}, owing to  the core-shift effect,
the core at 2.3\,GHz is expected 
to lie at a larger distance from the system center than at 8.4\,GHz.
However,  because of the  large scattering  
a proper astrometric analysis at 2.3\,GHz is  impossible.
Future observations should be done at high frequencies.
By measuring  core positions at many frequencies in the AGN M87,
 it was possible to determine the frequency dependence 
of the shift and find the converging location of all  observing-wavelength dependent cores. 
The six frequency observation of  M87  proved that
$\nu_{break_1}$ is at  43\,GHz  and that the  43 GHz-core is located  within 23 Schwarzschild radii 
from the system center  \citep{hada11}.
Similar observations for \lsi would be particularly interesting not only to prove 
whether   $\nu_{break_1}$  is in the infrared as expected
for microquasars, but in particular to test whether  the different cores at the different frequencies all trace ellipses with different sizes and  
all with the same period, 
as shown in the sketch of Fig.~\ref{sketch}. 
An accurate determination of the period is  mandatory to  establish the physical process responsible
for precession.
 \citet{massizimmermann10} computed the precessional period for the accretion disk in \lsi
under tidal forces of the Be star ($P_\mathrm{tidal-forces}$) and  under the effect
of frame dragging produced by the rotation of the compact object ($P_\mathrm{Lense-Thirring}$).
By using those equations,  we find  that $P_\mathrm{tidal-forces}$ of  28~d  would require 
the unrealistic value for the size of the accretion disk
of $R_\mathrm{out}=0.5-0.8\times 10^{13}$ cm, i.e, nearly the semimajor axis, and can therefore be ruled out.
In contrast, $P_\mathrm{Lense-Thirring}\simeq 28$\,d would be realistic  because it only implies
a slow rotator  with  dimensionless spin parameter 
of  $a_*=0.5 \times 10^{-3}$. 
However, whereas our study firmly rules out tidal forces, there could be  other processes than frame dragging  at work in \lsi, 
which require a more accurate determination of the precessional period to be investigated.
The orbital period is known with great accuracy and is P~=~26.4960 $\pm$ 0.0028~d \citep{gregory02},
whereas our determination of the precessional period indicates a period  of about 27-28~d. 
Therefore, it is important to establish, with future observations, whether  the two periods, orbital and precessional ones, are indeed similar but still different.
\begin{figure*}
\includegraphics[width=.9\textwidth]{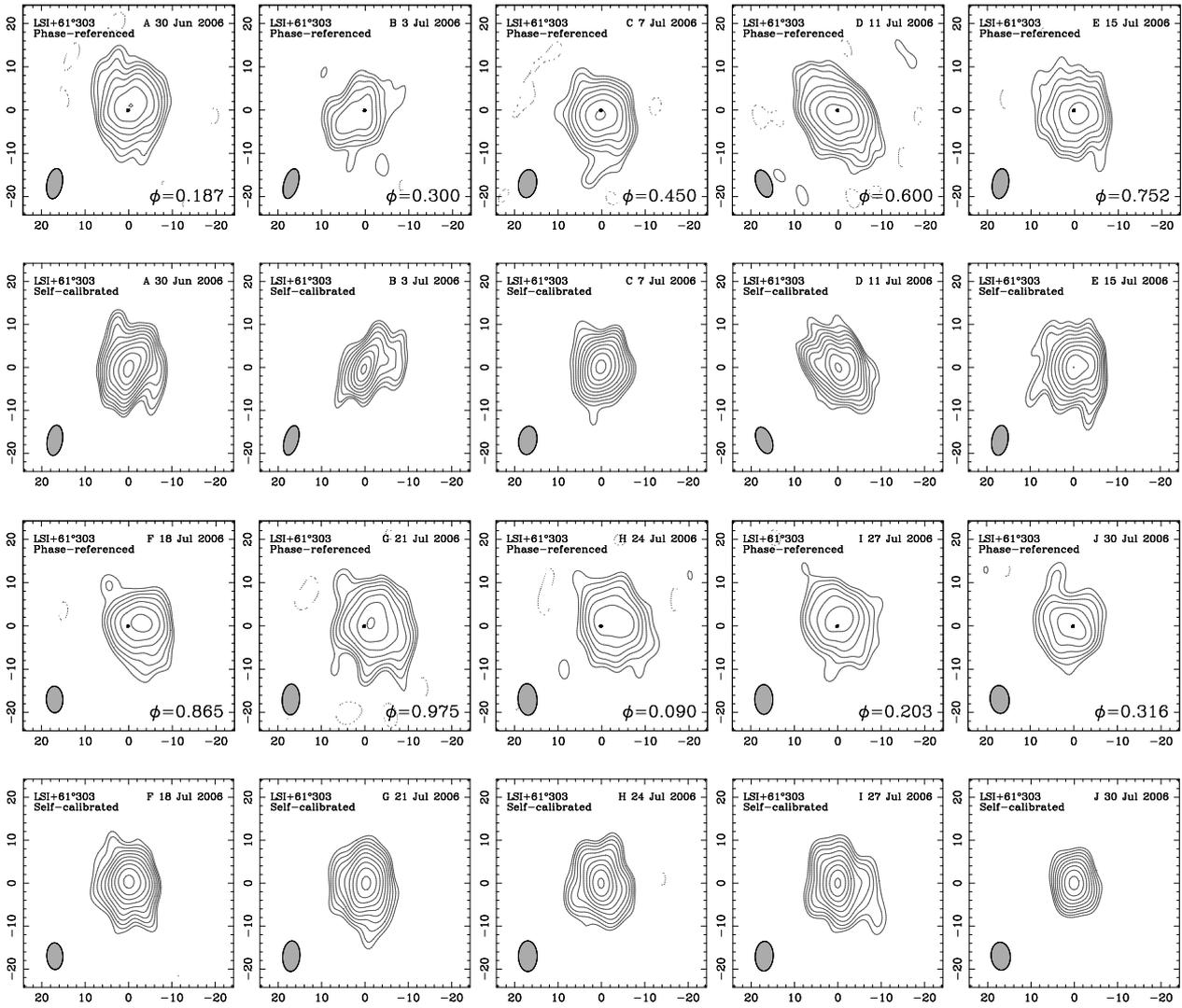}\\
         \label{maps13cm}
\caption{
Top: Images of VLBA runs A-J at 13 cm (2.3\,GHz) of \lsp.
For each run, two maps are presented, the phase referenced map and the
self-calibrated map. Both maps are
produced with uniform weight and the beam is  3.5 - 4.1 $\times$ 6.9 - 7.2.
Contour levels for all maps are -4, 4, 5.66, 8, 11.3, 16, 22.6, 32, 45.2,
64, 90.5, 128, 181 $\sigma $ 
(with 1$\sigma$=0.4 $-$ 0.6 mJy/beam for phase-referenced  maps, and 1$\sigma=$0.3 mJy/beam for
self-calibrated maps.)
}
\end{figure*}

\acknowledgements
We thank the referee for his  suggestions, which improved significantly the discussion
and 
 Lars Fuhrmann for  useful conversations and comments to the manuscript.
The Very Long Baseline Array is operated by the National Radio Astronomy Observatory, a facility of the National Science Foundation operated under cooperative agreement by Associated Universities, Inc. 
The Green Bank Interferometer is a facility of the National
Science Foundation operated by the NRAO in support of
NASA High Energy Astrophysics programs.
The work of L. Zimmermann is partly supported by the German Excellence
Initiative via the Bonn Cologne Graduate School. 
E.R. acknowledges partial support by the Spanish MICINN through grant AYA2009-13036-C02-02.
E.R. was supported by the COST action MP0905 "Black Holes in a Violent Universe" through the short-term scientific mission (STSM) MP0905-300711-008633.

\bibliographystyle{aa}

\end{document}